\documentclass[pra,twocolumn,showpacs,superscriptaddress]{revtex4}
\usepackage{amsmath}
\usepackage{graphicx}
\usepackage{amssymb}

\begin{document}

\title{Cavity optomechanical coupling assisted by an atomic gas}

\author{H. Ian}

\affiliation{Institute of Theoretical Physics, The Chinese Academy of Sciences,
Beijing, 100080, China}

\affiliation{Frontier Research System, The Institute of Physical and Chemical
Research (RIKEN), Wako-shi, Saitama 351-0198, Japan}

\author{Z. R. Gong}

\affiliation{Institute of Theoretical Physics, The Chinese Academy of Sciences,
Beijing, 100080, China}

\affiliation{Frontier Research System, The Institute of Physical and Chemical
Research (RIKEN), Wako-shi, Saitama 351-0198, Japan}

\author{Yu-xi Liu}

\affiliation{Frontier Research System, The Institute of Physical and Chemical
Research (RIKEN), Wako-shi, Saitama 351-0198, Japan}

\affiliation{CREST, Japan Science and Technology Agency (JST), Kawaguchi, Saitama
332-0012, Japan}

\author{C. P. Sun}

\affiliation{Institute of Theoretical Physics, The Chinese Academy of Sciences,
Beijing, 100080, China}

\affiliation{Frontier Research System, The Institute of Physical and Chemical
Research (RIKEN), Wako-shi, Saitama 351-0198, Japan}

\author{Franco Nori}

\affiliation{Frontier Research System, The Institute of Physical and Chemical
Research (RIKEN), Wako-shi, Saitama 351-0198, Japan}

\affiliation{CREST, Japan Science and Technology Agency (JST), Kawaguchi, Saitama
332-0012, Japan}

\affiliation{Center for Theoretical Physics, Physics Department, Center for the
Study of Complex Systems, The University of Michigan, Ann Arbor, Michigan
48109-1040, USA}

\date{\today}

\begin{abstract}
We theoretically study a cavity filled with atoms, which provides
the optical-mechanical interaction between the modified cavity
photonic field and a movable mirror at one end. We show that the
cavity field ``dresses'' these atoms, producing two types of
polaritons, effectively enhancing the radiation pressure of the
cavity field upon the end mirror, as well as establishing an
additional squeezing mode of the end mirror. This squeezing
produces an adiabatic entanglement, which is absent in usual
vacuum cavities, between the oscillating mirror and the rest of
the system. We analyze the entanglement and quantify it using the
Loschmidt echo and fidelity.
\end{abstract}


\maketitle

\section{Introduction}

Fabry-P\'{e}rot cavities (see, e.g., Ref.~\cite{metzger04,gigan06,kleckner06}) 
can trap incident light and
have a fixed mirror at one end and a movable mirror \cite{mirror}
at the other end. This movable mirror is allowed to oscillate
harmonically around a fixed position. The oscillating mirror
allows infinitesimal contractions and dilations of the cavity
length, resulting in a radiation pressure on the mirror which is
proportional to the intensity of the trapped cavity field. This
mechanism facilitates an optical-mechanical coupling between the
cavity field and the mirror and is now generating considerable
interest. In recent years, for example, a high-precision
spectrometer for detecting gravitational
waves~\cite{meers89,aguirre87} and an interferometric measurement
apparatus~\cite{tittonen99,arcizet06} have used movable cavity
mirrors as sensing devices. For detecting weak signals, a number
of experiments have reduced the thermal fluctuations in the
mirrors, effectively lowering the temperature of the
mirror~\cite{metzger04,gigan06,kleckner06}.

A key variable in previous designs is the number of photons
trapped inside the cavity. Since the radiation pressure on the
mirror is proportional to the photon number, it is desirable to
increase this photon number in order to increase the magnitude of
the radiation pressure and hence to control or cool down the
mirror more efficiently. Moreover, the cooling of a nanomechanical
resonator or an oscillating mirror is extensively studied recently
(e.g., in Refs.~\cite{fxue07-1,jqyou08}). This then naturally leads us to
conceive a cavity filled by a dielectric medium to achieve this
purpose. Specifically, following our previous idea in
Ref.~\cite{he07}, we now propose that this medium can be made of a
gas of two-level atoms.

Recently, Ref.~\cite{meiser06} has proposed a similar scheme to
target an interesting optical effect: the cavity mode forms an
optical lattice inside the cavity and arranges the free atoms that
were deposited into the cavity to form a Mott-insulator-like
medium with atoms trapped at the lattice sites. It was shown that,
with the atoms assuming an initial Bose-Einstein condensate
distribution, such an atomic condensate would act effectively as a
semi-transparent mirror itself and shift the cavity to function in
its ``superstrong coupling regime''. Nonetheless, based on
Monte-Carlo simulations~\cite{asboth07,meiser07}, there exist
disputes for the realizability of this proposal.

In this paper, we analyze the dynamical effect that occurs when
placing an atomic medium into a Fabry-P\'{e}rot cavity, but assuming
that the atoms have been placed inside a transparent gas chamber.
Due to the strengthened coupling, now enhanced by the mediating
atoms between the cavity field and the mirror, the resulting
three-component system (the gas of atoms, the cavity field, and
the oscillating mirror) induces interesting phenomena worth
investigating. We point out here that, in contrast with the BEC
atoms in Ref.~\cite{meiser06} that can only be realized at very
low temperatures, our gas of atoms makes use of low-energy
collective excitations, which avoids the stringent low-temperature
requirement.

To better extract the physical features of each part of this three-component
system, we assume adiabatic processes over different time scales.
We employ the Born-Oppenheimer approximation to study the dynamic
behavior of a micro-mirror by assuming it is a slow-varying part.
We also study the dynamic behavior of the atomic excitations as a
fast-varying process, while the reflected radiations from the mirror
stays relatively constant. The complex interactions between the system
components leads us to expect many interesting physical phenomena
including: (i) realizing an adiabatic entanglement process~\cite{suncp00-adia},
(ii) producing squeezed modes as in optical parametric oscillators,
(iii) detecting polaritons through the mechanical mode of the mirror,
and (iv) detecting the mechanical mode of the mirror through the polariton
spectrum.

We first describe the model in Sec.~\ref{sec:model}. The resulting
entanglement process is then described in Sec.~\ref{sec:entanglement}
and its quantification follows in Sec.~\ref{sec:quan_entanglement}.
The squeezed variance is derived in Sec.~\ref{sec:quad_variance}
and conclusions are presented in Sec.~\ref{sec:conclusion}.

\section{\label{sec:model}Atomic Optomechanics}

\subsection{The Exciton Model}

\begin{figure}
\includegraphics[bb=100bp 220bp 640bp 740bp,clip,width=2.8in]{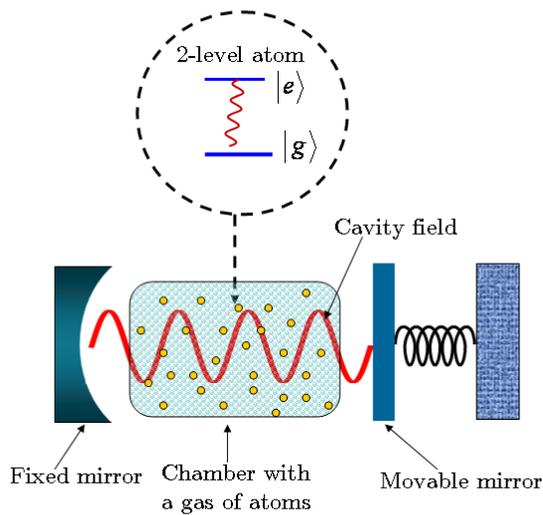}

\caption{(color online) Schematic diagram illustrating the system
with three main components: (i) a gas of two-level atoms, (ii) a
movable mirror at one end, and (iii) a cavity field mediating the
interaction between the atoms and the oscillating
mirror.\label{fig:model_setup}}
\end{figure}

As shown in Fig.~\ref{fig:model_setup}, the system we study here
consists of a gas of two-level atoms, each with the same eigen-frequency
$\Omega_{0}$, a Fabry-P\'{e}rot cavity carrying a photonic field with
mode frequency $\Omega_{\mathrm{C}}$, as well as a harmonically bounded
micro-mirror with coordinate $x$, momentum $p$, mass $m$ and oscillating
frequency $\Omega_{\mathrm{M}}$. The system Hamiltonian, with units
normalized according to $\hbar=1$ to simplify the notation, is
\begin{eqnarray}
H & = & \Omega_{0}\sum_{j}\sigma_{j}^{z}+\Omega_{\mathrm{C}}a^{\dagger}a+\sum_{j}(g_{j}\sigma_{j}^{+}a+g_{j}^{\ast}\sigma_{j}^{-}a^{\dagger})\nonumber \\
 &  & +\frac{p^{2}}{2m}+\frac{1}{2}m\Omega_{\mathrm{M}}^{2}x^{2}+\eta a^{\dagger}ax.\label{eq:orig_tot_ham}
\end{eqnarray}
In Eq.~(\ref{eq:orig_tot_ham}), the Pauli matrix
$\sigma_{j}^{z}=\left|e_j\right\rangle \left\langle e_j\right|$
denotes the internal energy of each two-level atom, while
$\sigma_{j}^{+}=\left|e_j\right\rangle \left\langle g_j\right|$ and
$\sigma_{j}^{-}=\left|g_j\right\rangle \left\langle e_j\right|$ in the
last term of the first line denote the flip-up and flip-down
operators of the $j$-th atom. Here, $a$ and $a^{\dagger}$ denote,
respectively, the annihilation and creation operators of the
cavity field. The last term of the second line is a
radiation-pressure-type interaction on the mirror, which is
proportional to the incident photon number. We assume that no
direct interaction exists between the atoms and the mirror; the
indirect interaction between them solely relies on the cavity
field.

Since all the atoms have the same frequency $\Omega_{0}$, we can
consider the gas of atoms as a whole to be a
Hopfield-dielectric~\cite{hopfield58} filling the cavity, and its
behavior described by collective low-energy excitations using the
exciton annihilation operator~\cite{he07}
\begin{equation}
b=\lim_{N\to\infty}\sum_{j}^{N}\frac{g_{j}^{\ast}}{G}\sigma_{j}^{-}\end{equation}
 and its Hermitian conjugate $b^{\dagger}$, where \[
G=\sqrt{\sum_{j=1}^{N}|g_{j}|^{2}}\]
 can be understood as the total coupling strength. The exciton operators
$b$ and $b^{\dagger}$ are bosonic and consistent with the Dicke
model; a similar spin-bosonization technique has been used to study
nuclear spins~\cite{song05}. The resulting Hamiltonian for the
system can then be written as \begin{eqnarray}
H & = & \Omega_{0}b^{\dagger}b+(\Omega_{\mathrm{C}}+\eta x)a^{\dagger}a+G(b^{\dagger}a+ba^{\dagger})\nonumber \\
 &  & +\frac{p^{2}}{2m}+\frac{1}{2}m\Omega_{\mathrm{M}}^{2}x^{2}.\label{eq:tot_ham}\end{eqnarray}

\subsection{Interaction between the oscillating mirror and the polaritons}

The coupling between the excitons and the cavity field can lead to
the emergence of \emph{dressed excitons}, here denoted as
polaritons. In the adiabatic limit of the oscillating mirror, that
is, when the mirror coordinate $x$ stays unchanged with respect to
the fast-varying field occupation number $a^{\dagger}a$, we can
diagonalize the interaction between the excitons and the cavity
field by rotating the Hilbert space of these two components
through an angle\begin{equation}
\theta=\arctan\left(\frac{2G}{\Omega_{0}-\Omega_{\mathrm{C}}-\eta
x}\right)\end{equation}
 for which we define a unitary transformation\begin{eqnarray}
A & = & a\,\cos\negmedspace\left(\frac{\theta}{2}\right)-b\,\sin\negmedspace\left(\frac{\theta}{2}\right),\\
B & = & a\,\sin\negmedspace\left(\frac{\theta}{2}\right)+b\,\cos\negmedspace\left(\frac{\theta}{2}\right).\end{eqnarray}

The $A$ and $B$ operators above still obey bosonic commutation
relations and can be understood as ``dressed exciton modes'' that
mix atomic excitations $b$ with the cavity field $a$. In other
words, these dressed exciton modes are polaritons~\cite{he07} of a
phonon mode $A$ and an optical mode $B$.

Under this view, the Hamiltonian of our system in Eq.~(\ref{eq:tot_ham}) can be divided
into two portions, the Hamiltonian $H_{\mathrm{M}}$ of the
mirror's oscillation and the potential $V$ from the polaritons
acting on the mirror, i.e.\begin{eqnarray}
H & = & H_{\mathrm{M}}+V,\\
H_{\mathrm{M}} & = & H_{\mathrm{mirror\ oscillations}}\\
 & = & \frac{p^{2}}{2m}+\frac{1}{2}m\Omega_{\mathrm{M}}^{2}x^{2},\\
V & = & V_{\mathrm{polaritons\ on\ mirror}}\\
 & = & \frac{1}{2}(\Omega_{0}+\Omega_{\mathrm{C}}+\eta x)(A^{\dagger}A+B^{\dagger}B)\label{eq:polariton_pot}\\
 &  & -\frac{1}{2}\sqrt{(\Omega_{0}-\Omega_{\mathrm{C}}-\eta x)^{2}+4G^{2}}(A^{\dagger}A-B^{\dagger}B).\nonumber \end{eqnarray}
 The potential $V$ in Eq.~(\ref{eq:polariton_pot}) quantifies the
interaction between the mechanical mirror and the modes of the cavity.
Without the {}``filling'' atoms, the cavity mode is simply the photon
field $a$, and this potential $V$ will degenerate back to a linear
radiation pressure impinging on the mirror, if we do not consider
the nonlinear Kerr effect that could be induced by the wave detuning
due to the flexible length of the cavity~\cite{mancini94,zrgong}.

The atoms let the linear radiation pressure be proportional to the
total number ($A^{\dagger}A+B^{\dagger}B$) of polaritons [$\eta
x(A^{\dagger}A+B^{\dagger}B)$ in Eq.~(\ref{eq:polariton_pot})]
rather than the number ($a^{\dagger}a$) of photons [$\eta
a^{\dagger}ax$ in Eq.~(\ref{eq:orig_tot_ham})]. Moreover, the
atoms also impose an additional nonlinear term (the second term in
Eq.~(\ref{eq:polariton_pot})) for non-zero coupling constant $G$.

Note that this nonlinear effect, in the second term of
Eq.~(\ref{eq:polariton_pot}), increases when increasing the number
$N$ of filled atoms because $G$ grows with $N$. Thus, the gas of
atoms enhances the coupling between the cavity field and the mirror.
This enhanced coupling would produce squeezed states of the mirror
mode and also entanglement between the mirror and the polaritons,
which will be discussed in Sec.~\ref{sec:entanglement}. Without
the intervening atoms, the potential $V$ simply introduces a
displacement to the mirror, producing neither squeezing nor
entanglement.

\section{\label{sec:entanglement}Adiabatic Entanglement and Evolution under
Squeezing}

\subsection{\label{sub:geo_entanglement}Entanglement Using the Born-Oppenheimer
Approximation}

By considering fast-varying polariton modes and slow-varying mirror
modes, we can write the wave vector at time $t$ for our system under
the Born-Oppenheimer approximation \begin{equation}
\left|\psi(t)\right\rangle =\sum_{\mathbf{n}}\left|\mathbf{n}\right\rangle \otimes\left|\phi(\mathbf{n},t)\right\rangle \label{eq:state_vec}\end{equation}
 where $\mathbf{n}=\left\{ n_{A},n_{B}\right\} $ denotes the collective
index of energy levels of the polariton modes $A$ and $B$. Thus,
$A^{\dagger}A\left|\mathbf{n}\right\rangle =n_{A}\left|\mathbf{n}\right\rangle $
and $B^{\dagger}B\left|\mathbf{n}\right\rangle =n_{B}\left|\mathbf{n}\right\rangle $.
Here, $\left|\mathbf{n}\right\rangle $ describes the time-independent
wave vector for the polariton space in its adiabatic limit and $\left|\phi(\mathbf{n},t)\right\rangle $
the time-dependent wave vector for the mirror. The potential $V$
in Eq.~(\ref{eq:polariton_pot}) then becomes an effective c-number
according to the eigenspectrum $\mathbf{n}$,\begin{eqnarray}
V_{\mathbf{n}} & = & \frac{1}{2}(\Omega_{0}+\Omega_{\mathrm{C}}+\eta x)(n_{B}+n_{A})\nonumber \\
 &  & +\frac{1}{2}\sqrt{(\Omega_{0}-\Omega_{\mathrm{C}}-\eta x)^{2}+4G^{2}}(n_{B}-n_{A}).\label{eq:BO_pot}\end{eqnarray}

We consider the displacement of the mirror $x$ to be small around
its equilibrium position $x=0$ and thus expand
Eq.~(\ref{eq:BO_pot}) up to second order in $x$
\begin{multline}
V_{\mathbf{n}}=\frac{1}{2}(\Omega_{0}+\Omega_{\mathrm{C}})(n_{B}+n_{A})\\
+\frac{1}{2}\sqrt{(\Omega_{0}-\Omega_{\mathrm{C}})^{2}+4G^{2}}(n_{B}-n_{A})\\
+\frac{\eta}{2}\left[(n_{B}+n_{A})-\frac{(\Omega_{0}-\Omega_{\mathrm{C}})(n_{B}-n_{A})}{\sqrt{(\Omega_{0}-\Omega_{\mathrm{C}})^{2}+4G^{2}}}\right]x\\
+\frac{N|g|^{2}\eta^{2}(n_{B}-n_{A})}{\left((\Omega_{0}-\Omega_{\mathrm{C}})^{2}+4G^{2}\right)^{\frac{3}{2}}}x^{2}.\label{eq:eff_pot}
\end{multline}
Using Eq.~(\ref{eq:eff_pot}), when the polariton modes are in
state $\left|\mathbf{n}\right\rangle $, the effective Hamiltonian
operating on the mirror is
\begin{equation} H_{\mathbf{n}}=H_{\mathrm{M}}
+V_{\mathbf{n}}.\label{eq:eff_ham}
\end{equation}
If we prepare an initial state of the system
\begin{equation}
\left|\psi(0)\right\rangle
=\sum_{\mathbf{n}}\lambda_{\mathbf{n}}\left|\mathbf{n}\right\rangle
\otimes\left|\phi(0)\right\rangle
\end{equation}
 where $\lambda_{\mathbf{n}}$ is the expansion coefficient, then
the mirror wave subvector will evolve along the path generated by
the effective Hamiltonian $H_{\mathbf{n}}$ \begin{eqnarray}
\left|\psi(t)\right\rangle  & = & \sum_{\mathbf{n}}\lambda_{\mathbf{n}}\left|\mathbf{n}\right\rangle \otimes\left|\phi_{\mathbf{n}}(t)\right\rangle ,\\
\left|\phi_{\mathbf{n}}(t)\right\rangle  & = & e^{-iH_{\mathbf{n}}t}\left|\phi(0)\right\rangle .\label{eq:mir_vec}\end{eqnarray}
 In other words, the final state of the mirror is determined by or
dependent on the state of the polaritons in their adiabatic limit;
specifically, the number distribution of the polaritons $\left\{ n_{A},n_{B}\right\} $
will decide the evolution of the mirror.

Geometrically, if the initial state $\left|\phi(0)\right\rangle $
was conceived to be represented by a point on a manifold over the
Hilbert space of the mirror, then the effective Hamiltonians
$H_{\mathbf{n}}$ and $H_{\mathbf{m}}$ for
$\mathbf{n}\neq\mathbf{m}$ can be regarded as generators of the
motion of the same vector $\left|\phi(0)\right\rangle $ towards
different directions over the manifold. The evolution over time
due to different generators will leave trajectories of different
branches of paths on the manifold. The end points
$\left|\phi_{\mathbf{n}}(t)\right\rangle $ and
$\left|\phi_{\mathbf{m}}(t)\right\rangle $ of the paths are
separated and the separation depends on the discrepancy between
$H_{\mathbf{n}}$ and $H_{\mathbf{m}}$ induced by different
polariton distributions. The nonzero separation reflects
geometrically the adiabatic entanglement of the mirror and the
polaritons.

The original concept of adiabatic entanglement proposed in Ref.~\cite{suncp00-adia}
concerns a two-component model of a fast-varying main system to be
measured and a slow-varying detector apparatus. The system and the
detector entangles over time as described by Eq.~(\ref{eq:state_vec}).
We hence regard the triplet system discussed above as a practical
realization of the adiabatic entanglement model by comparing the polaritons
with the system and the mirror with the detector.

\subsection{Evolution of squeezed coherent states of the mirror}

Before quantifying the entanglement described above, we first study
the dynamics of the mirror via the effective Hamiltonian $H_{\mathbf{n}}$.
If we write the coordinate operator $x$ and the momentum operator
$p$ of the mirror in their creation and annihilation operator form
\begin{eqnarray}
x & = & \frac{1}{\sqrt{2m\Omega_{\mathrm{M}}}}(c+c^{\dagger}),\label{eq:cordinate}\\
p & = & -i\sqrt{\frac{m\Omega_{\mathrm{M}}}{2}}(c-c^{\dagger}),
\end{eqnarray}
the effective Hamiltonian, i.e. Eq.~(\ref{eq:eff_ham}), then
reads
\begin{equation}
H_{\mathbf{n}}=(\Omega_{\mathrm{M}}+2\alpha_{\mathbf{n}})c^{\dagger}c+\alpha_{\mathbf{n}}(c^{2}+c^{\dagger2})+\beta_{\mathbf{n}}(c+c^{\dagger})+\gamma_{\mathbf{n}},\label{eq:quantized_ham}
\end{equation}
 where the coefficients depend on the polariton modes
 \begin{align}
\alpha_{\mathbf{n}}= & \frac{G^{2}\eta^{2}(n_{B}-n_{A})}{2m\Omega_{\mathrm{M}}\left((\Omega_{0}-\Omega_{\mathrm{C}})^{2}+4G^{2}\right)^{\frac{3}{2}}},\label{eq:alpha}\\
\beta_{\mathbf{n}}= & \frac{\eta}{\sqrt{8m\Omega_{\mathrm{M}}}}\left[(n_{B}+n_{A})-\frac{(\Omega_{0}-\Omega_{\mathrm{C}})(n_{B}-n_{A})}{\sqrt{(\Omega_{0}-\Omega_{\mathrm{C}})^{2}+4G^{2}}}\right],\label{eq:beta}\\
\gamma_{\mathbf{n}}= & \frac{1}{2}(\Omega_{0}+\Omega_{\mathrm{C}})(n_{B}+n_{A})\\
 & +\frac{1}{2}\sqrt{(\Omega_{0}-\Omega_{\mathrm{C}})^{2}+4G^{2}}\;(n_{B}-n_{A})\nonumber \\
 & +\frac{N|g|^{2}\eta^{2}(n_{B}-n_{A})}{m\Omega_{\mathrm{M}}\left((\Omega_{0}-\Omega_{\mathrm{C}})^{2}+4G^{2}\right)^{\frac{2}{3}}}.\nonumber
 \end{align}

The first- and second-order terms of $c$ and $c^{\dagger}$ in Eq.~(\ref{eq:quantized_ham})
can be recognized as the polaritons inducing a squeezed coherent state
in the mirror. The amount of displacement can be found by writing
Eq.~(\ref{eq:quantized_ham}) as
\begin{align}
H_{\mathbf{n}}= & D^{\dagger}\negmedspace\left(\frac{\beta_{\mathbf{n}}}{\Omega_{\mathrm{M}}+4\alpha_{\mathbf{n}}}\right)H'_{\mathbf{n}}\; D\negmedspace\left(\frac{\beta_{\mathbf{n}}}{\Omega_{\mathrm{M}}+4\alpha_{\mathbf{n}}}\right)
\end{align}
 where $D(\alpha)=\exp\left\{ \alpha^{\ast}a-\alpha a^{\dagger}\right\} $
is the displacement operator. The resulting Hamiltonian in the displaced
space is \begin{equation}
H'_{\mathbf{n}}=(\Omega_{\mathrm{M}}+2\alpha_{\mathbf{n}})c^{\dagger}c+\alpha_{\mathbf{n}}(c^{2}+c^{\dagger2})-\frac{\beta_{\mathbf{n}}^{2}}{\Omega_{\mathrm{M}}+4\alpha_{\mathbf{n}}}+\gamma_{\mathbf{n}}.\label{eq:dis_ham}\end{equation}
 The amount of squeezing can be found by further diagonalizing Eq.~(\ref{eq:dis_ham})
through a Bogoliubov transformation\begin{eqnarray}
C_{\mathbf{n}} & = & \mu_{\mathbf{n}}\,c-\nu_{\mathbf{n}}\,c^{\dagger},\label{eq:bogo_tfm}\\
\mu_{\mathbf{n}} & = & \frac{1}{2}\left[\left(\frac{\Omega_{\mathrm{M}}}{\Omega_{\mathrm{M}}+4\alpha_{\mathbf{n}}}\right)^{\frac{1}{4}}+\left(\frac{\Omega_{\mathrm{M}}+4\alpha_{\mathbf{n}}}{\Omega_{\mathrm{M}}}\right)^{\frac{1}{4}}\right],\\
\nu_{\mathbf{n}} & = & \frac{1}{2}\left[\left(\frac{\Omega_{\mathrm{M}}}{\Omega_{\mathrm{M}}+4\alpha_{\mathbf{n}}}\right)^{\frac{1}{4}}-\left(\frac{\Omega_{\mathrm{M}}+4\alpha_{\mathbf{n}}}{\Omega_{\mathrm{M}}}\right)^{\frac{1}{4}}\right],\end{eqnarray}
 for which the resulting Hamiltonian becomes\begin{equation}
H'_{\mathbf{n}}=\Omega_{\mathrm{M},\mathbf{n}}\:
C_{\mathbf{n}}^{\dagger}C_{\mathbf{n}}+\zeta_{\mathbf{n}}\label{eq:linear_ham}\end{equation}
where $\Omega_{\mathrm{M},\mathbf{n}}$ denotes the modified
pseudo-energy splitting of the transformed mirror excitations
according to $C_{\mathbf{n}}$ and $C_{\mathbf{n}}^{\dagger}$
\begin{equation}
\Omega_{\mathrm{M},\mathbf{n}}=\sqrt{\Omega_{\mathrm{M}}(\Omega_{\mathrm{M}}+4\alpha_{\mathbf{n}})},
\end{equation}
 and $\zeta_{\mathbf{n}}$ denotes the non-operator terms
\begin{equation}
\zeta_{\mathbf{n}}=-\,\frac{\left(\sqrt{\Omega_{\mathrm{M}}}-\sqrt{\Omega_{\mathrm{M}}+4\alpha_{\mathbf{n}}}\right)^{2}}{4}-\frac{\beta_{\mathbf{n}}^{2}}{\Omega_{\mathrm{M}}+4\alpha_{\mathbf{n}}}+\gamma_{\mathbf{n}}.
\end{equation}
Here, $\Omega_{\mathrm{M},\mathbf{n}}$ is called a
pseudo-frequency because it might become imaginary for some cases
of the index $\mathbf{n}$. This reflects the fact that the
distribution of the polaritons have a strong influence over the
time evolution of the mirror, as we have pointed out above. In the
next subsection, we shall give more definite consideration for
$\Omega_{\mathrm{M},\mathbf{n}}$ being real or imaginary when
discussing the Loschmidt echo.

The transformation Eq.~(\ref{eq:bogo_tfm}) is physically equivalent
to squeezing the operator $c$. To simplify the derivation we shall
develop in the following, we define this squeezing process inversely
with the operator $S_{\mathbf{n}}$
\begin{eqnarray}
c & = & S_{\mathbf{n}}^{\dagger}\; C_{\mathbf{n}}\; S_{\mathbf{n}}\\
S_{\mathbf{n}} & = & \exp\left\{ \frac{r_{\mathbf{n}}}{2}\, C_{\mathbf{n}}^{2}-\frac{r_{\mathbf{n}}}{2}\, C_{\mathbf{n}}^{\dagger2}\right\} \label{eq:squeeze_opt}
\end{eqnarray}
where $\cosh r_{\mathbf{n}}=\mu_{\mathbf{n}}$. Over an initial
coherent state $\left|\alpha\right\rangle $ with
$c\left|\alpha\right\rangle =\alpha\left|\alpha\right\rangle $, we
can define a special {}``coherent state''
\begin{equation}
\left|\alpha\right\rangle
_{\mathbf{n}}=S_{\mathbf{n}}\left|\alpha\right\rangle
\end{equation}
according to the operator $C_{\mathbf{n}}$, i.e. \[
C_{\mathbf{n}}\left|\alpha\right\rangle
_{\mathbf{n}}=\alpha\left|\alpha\right\rangle _{\mathbf{n}}.\] The
time evolution of the mirror, starting from an initial vacuum
state, can then be computed as:
\begin{eqnarray}
\left|\phi_{\mathbf{n}}(t)\right\rangle  & = & e^{-iH_{\mathbf{n}}t}\left|0\right\rangle \nonumber \\
 & = & D^{\dagger}\negmedspace\left(\frac{\beta_{\mathbf{n}}}{\Omega_{\mathrm{M}}+4\alpha_{\mathbf{n}}}\right)e^{-iH'_{\mathbf{n}}t}\left|\frac{\beta_{\mathbf{n}}}{\Omega_{\mathrm{M}}+4\alpha_{\mathbf{n}}}\right\rangle \nonumber \\
 & = & D^{\dagger}\negmedspace\left(\frac{\beta_{\mathbf{n}}}{\Omega_{\mathrm{M}}+4\alpha_{\mathbf{n}}}\right)S_{\mathbf{n}}^{\dagger}(t)\left|\frac{\beta_{\mathbf{n}}}{\Omega_{\mathrm{M}}+4\alpha_{\mathbf{n}}}\right\rangle _{\mathbf{n}}\nonumber \\
 & = & \left|\frac{\beta_{\mathbf{n}}}{\Omega_{\mathrm{M}}+4\alpha_{\mathbf{n}}}(e^{-i\Omega_{\mathrm{M},\mathbf{n}}t}-1)\right\rangle \label{eq:final_state}
\end{eqnarray}
where $S_{\mathbf{n}}^{\dagger}$ is the Hermitian conjugate of the
squeezing operator $S_{\mathbf{n}}$ in Eq.~(\ref{eq:squeeze_opt})
in the Heisenberg picture; more explicitly,\begin{equation}
S_{\mathbf{n}}^{\dagger}(t)=\exp\left\{
\frac{r_{\mathbf{n}}(t)}{2}C_{\mathbf{n}}^{\dagger2}-\frac{r_{\mathbf{n}}(t)}{2}C_{\mathbf{n}}^{2}\right\}
\end{equation}
with $r_{\mathbf{n}}(t)=r_{\mathbf{n}}\exp\left\{ -i\Omega_{\mathrm{M},
\mathbf{n}}t\right\} $. This derivation is similar to the
technique used in Ref.~\cite{ydwang04} for computing the evolution
of squeezed states. The difference is that the coupling nature of
the system we consider permits the entanglement of the oscillating
mirror even when initialized from a vacuum state, which avoids the
difficulty of preparing a coherent superposition. The squeezed
states using polaritons have been studied in Ref.~\cite{xdhu96-1},
while phonon squeezed states in Ref.~\cite{xdhu96-2}

\section{\label{sec:quan_entanglement}Quantification of Decoherence and Entanglement}

\subsection{Loschmidt Echo}

At the end of Sec.~\ref{sub:geo_entanglement}, we interpreted the
adiabatic entanglement as two distinct end points of the evolution
over a manifold. The metric distance between the two points naturally
becomes an appropriate measure of the degree of coherence or correlation
between the two quantum states. The Loschmidt echo, which has been
known to characterize the decoherence of a perturbed system~\cite{cucchietti03},
plays the role of metric. Originally this echo was defined as the
wave function overlap between the states with and without the presence
of the perturbing potential. This echo exactly describes the dynamic
sensitivity of the system in the context of quantum chaos.

In our case, the perturbation potential can be understood as $(H_{\mathbf{n}}-H_{\mathbf{m}})$
and the echo as \[
L_{\mathbf{n,m}}(t)=|\left\langle \phi_{\mathbf{n}}(t)|\phi_{\mathbf{m}}(t)\right\rangle |.\]
 Using Eq.~(\ref{eq:final_state}), we find
\begin{multline}
L_{\mathbf{n,m}}(t) = 
\exp\left\{-\sum_{i=\mathbf{m},\mathbf{n}}\frac{2\beta_{i}^{2}}
{(\Omega_{\mathrm{M}}+4\alpha_{i})^{2}}\sin^{2}
\left(\frac{\Omega_{\mathrm{M},i}}{2}t\right)\right.\nonumber\\
+\left[\sum_{i=\mathbf{m},\mathbf{n}}\sin^{2}\left(\frac{\Omega_{\mathrm{M},i}}{2}t\right)
-\sin^{2}\left(\Omega_{\mathrm{M},\mathbf{n-m}}\,t\right)\right]
\times\nonumber\\
\left.\frac{\beta_{\mathbf{n}}\beta_{\mathbf{m}}}{(\Omega_{\mathrm{M}}+4\alpha_{\mathbf{n}})
(\Omega_{\mathrm{M}}+4\alpha_{\mathbf{m}})}\right\}.\label{eq:Loschmidt}
\end{multline}
with
\begin{equation}
\Omega_{\mathrm{M},\mathbf{n-m}}=\frac{1}{2}(\Omega_{\mathrm{M},\mathbf{n}}-\Omega_{\mathrm{M},\mathbf{m}}).
\end{equation}
Note that when $\Omega_{\mathrm{M},\mathbf{n}}$ and
$\Omega_{\mathrm{M},\mathbf{m}}$ are real, the echo exhibits a
cycling collapse and revival, similar to the decoherence effect
shown in Ref.~\cite{ydwang04}, only that the oscillation is not
simply sinusoidal. When the two pseudo-frequencies are indeed
imaginary, which occurs when
\begin{equation}
\alpha_{\mathbf{n}}<-\,\frac{\Omega_{\mathrm{C}}}{4}
\end{equation}
 for some $\mathbf{n}$ or equivalently\begin{equation}
\left(n_{A}-n_{B}\right)\;>\frac{m\Omega_{\mathrm{M}}^{2}((\Omega_{0}-\Omega_{\mathrm{C}})^{2}+4G^{2})^{\frac{3}{2}}}{2G^{2}\eta^{2}},\label{eq:critical_cond}\end{equation}
 the sinusoidal functions will become hyperbolic and the echo will
damp with time exponentially. Whether the latter can happen depends
on the difference between the excitation numbers of the polaritons.
The requiring difference being large or small depends on the coupling
constant $G$, which in turn increases with the number $N$ of atoms
in the cavity. In other words, we can operate our system in two regimes:
for either periodic or hyperbolic Loschmidt echos, based on the number
$N$ of atoms.

\begin{figure}
\includegraphics[width=3.2in]{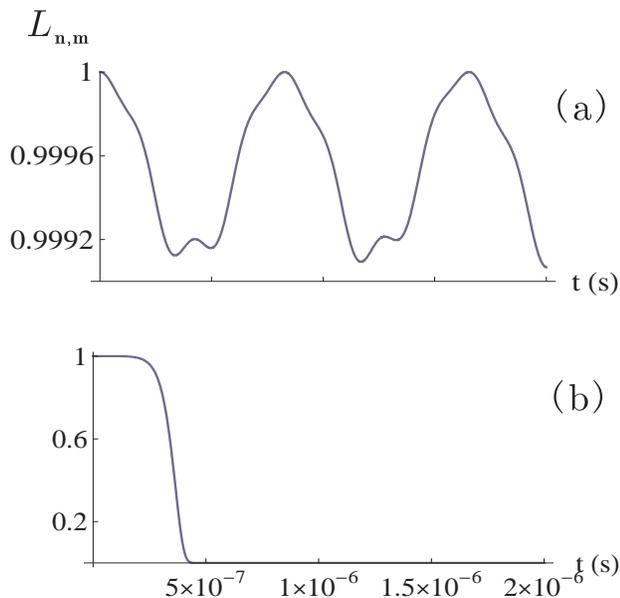}

\caption{Plots of the Loschmidt echo $L_{\mathbf{n,m}}$ for two mirror states
under the adiabatic entanglement for the operating regimes of (a)
circular functions and (b) hyperbolic functions. \label{fig:loschmidt_echo}}
\end{figure}

Figure~\ref{fig:loschmidt_echo} plots the echo $L_{\mathbf{n,m}}$
between two mirror states over the same period of time for the two
regimes. Without loss of generality, the parameters are all set to
orders of magnitude accessible to current experiments: $\Omega_{\mathrm{M}}/2\pi=10$
MHz, $\alpha_{\mathbf{n}}/2\pi=10^{11}$ Hz and $\beta_{\mathbf{n}}/2\pi=10^{7}$
Hz. The periodic Loschmidt echo in Fig.~\ref{fig:loschmidt_echo}(a)
demonstrates the collapse and revival of decoherence between two mirror
states whereas the hyperbolic type of echo in Fig.~\ref{fig:loschmidt_echo}(b)
shows a straight one-way decoherence. In the language of Ref.~\cite{cucchietti03},
\[
\Omega_{\mathrm{M}}+4\alpha_{\mathbf{n}}=0\]
 is a critical point of dynamic sensitity. When Eq.~(\ref{eq:critical_cond})
is met, the evolution paths of the two states on the manifold always
remain close to each other, giving an almost perfect echo. Once the
parameters cross into the opposite side of Eq.~(\ref{eq:critical_cond}),
the echo gets lost almost instantly with no comeback, as shown in
Fig.~\ref{fig:loschmidt_echo}(b).

\subsection{Fidelity}

\begin{figure}
\includegraphics[width=3in]{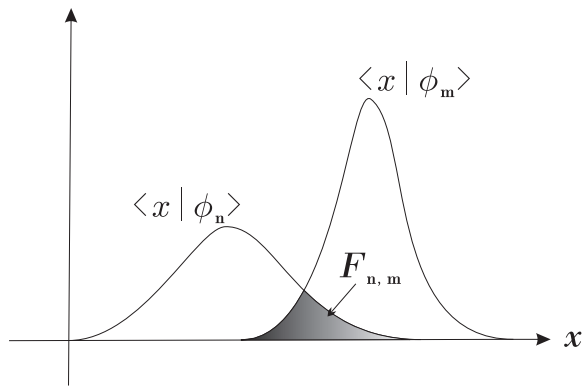}

\caption{Illustration of the fidelity between two quantum state vectors represented
in coordinate space as Gaussian functions\label{fig:fidelity}}
\end{figure}

Fidelity serves as another metric for measuring the correlation between
two quantum states. When seen in the coordinate space, the fidelity
roughly represents the overlap of the spatial wave packets of the
two states (illustrated in Fig.(\ref{fig:fidelity}) as the shaded
region). Defined as the inner product of the ground states of two
Hamiltonians, its physical meaning differs from that of the Loschmidt
echo in that it is not a time-dependent measure of the distance between
two evolving states, but a static estimate of the differentiating
effects of two dynamic evolution generators. The fidelity has recently
seen extensive applications on characterizing quantum phase transitions
in strong correlated systems~\cite{zanardi07}.

For our model, we use the fidelity to estimate the effects between
different polariton distributions on the mirror, under the Born-Oppenheimer
approximation. From Eqs.~(\ref{eq:dis_ham}), (\ref{eq:linear_ham})
and (\ref{eq:squeeze_opt}), the mirror ground state of the effective
Hamiltonian in the adiabatic limit is the squeezed coherent vacuum
state $\left|0\right\rangle _{\mathbf{n}}$ displaced by the amount
$\beta_{\mathbf{n}}/(\Omega_{\mathrm{M}}+4\alpha_{\mathbf{n}})$.
The fidelity, as the overlap of the ground states of two branching
Hamiltonians $H_{\mathbf{n}}$ and $H_{\mathbf{m}}$ ($\mathbf{n}\neq\mathbf{m}$),
can then be computed as the inner product of two coherent states\begin{align}
F_{\mathbf{n,m}} & =\left|\left\langle 0\right|S_{\mathbf{n}}^{\dagger}D^{\dagger}\negthickspace\left(\frac{\beta_{\mathbf{n}}}{\Omega_{\mathrm{M}}+4\alpha_{\mathbf{n}}}\right)\negmedspace D\negthickspace\left(\frac{\beta_{\mathbf{m}}}{\Omega_{\mathrm{M}}+4\alpha_{\mathbf{m}}}\right)\negthickspace S_{\mathbf{m}}\left|0\right\rangle \right|\nonumber \\
 & =\exp\left\{ -\frac{1}{2}\left(\frac{\beta_{\mathbf{n}}}{\Omega_{\mathrm{M}}+4\alpha_{\mathbf{n}}}-\frac{\beta_{\mathbf{m}}}{\Omega_{\mathrm{M}}+4\alpha_{\mathbf{m}}}\right)^{2}\right\} .\end{align}

\begin{figure}
\includegraphics[width=3.2in]{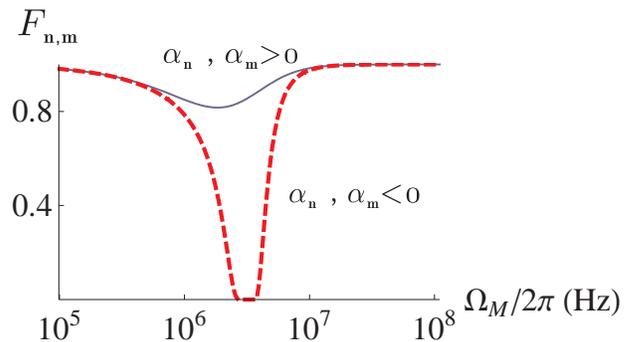}

\caption{(Color online) Semi-log plots of the fidelity $F_{\mathbf{n,m}}$
between two mirror states as a function of the mirror oscillating
frequency $\Omega_{\mathrm{M}}$ for two typical operating regimes.
The solid line represents the case for $\alpha_{\mathbf{n}},\alpha_{\mathbf{m}}>0$
while the dashed line for $\alpha_{\mathbf{n}},\alpha_{\mathbf{m}}<0$.
\label{fig:fidelity_plot}}
\end{figure}

We hence see that the wave-packet overlap $F_{\mathbf{n,m}}$ depends
on various parameters and generally, based on the relations of $\alpha_{\mathbf{n}}$
and $\beta_{\mathbf{n}}$ with $\Omega_{\mathrm{M}}$ (Cf. Eqs.~(\ref{eq:alpha})
and (\ref{eq:beta})), the overlap decreases for increasing $\Omega_{\mathrm{M}}$
over a normal mechanical oscillating frequency range. Figure \ref{fig:fidelity_plot}
shows the plot of the fidelity with the ordinate being the mirror
frequency over the range 100 kHz to 100 MHz in a logarithmic scale
for two typical values of the parameters: $\alpha_{\mathbf{n}},\alpha_{\mathbf{m}}>0$
and $\alpha_{\mathbf{n}},\alpha_{\mathbf{m}}<0$. The order of magnitude
of $\alpha_{\mathbf{n}}$ and $\beta_{\mathbf{n}}$ are set to ranges
consistent with the values given in the last subsection.

The low frequency range coincides with our expectation that a higher
oscillating frequency of the mirror will render itself more vulnerable
to the effect of the polaritons and induce its entanglement with other
system components faster. When $\Omega_{\mathrm{M}}$ further increases,
the different operating regimes studied using the Loschmidt echo manifest
themselves more apparently. For $\alpha_{\mathbf{n}},\alpha_{\mathbf{m}}>0$,
the two ground states of the mirror always remain close to each other,
corresponding to the periodic collapse and revival region for the
Loschmidt echo, and hence the fidelity retains its value close to
$1$. Whereas for $\alpha_{\mathbf{n}},\alpha_{\mathbf{m}}<0$, it
might cross into the hyperbolic operating region, where $\Omega_{\mathrm{M}}+4\alpha_{\mathbf{n}}<0$.
For the later case, the fidelity drops to $0$ near the critical point
$\Omega_{\mathrm{M}}+4\alpha_{\mathbf{n}}=0$, simulating the behavior of a
phase transition.

\section{\label{sec:quad_variance}Squeezed Quadrature Variance of the Mirror}

The gas of atoms inside the cavity also acts like an optical parametric
oscillator when regarded as a cavity dielectric. The original photon
field traveling in the cavity vacuum is dressed by the atoms into
two polariton modes. These two modes in their adiabatic limit act
on the mirror as if confining the mirror oscillation in a nonlinear
medium (Cf. Eqs.~(\ref{eq:polariton_pot}) and (\ref{eq:quantized_ham})
in which the polariton-mirror mode coupling is nonlinear). This case
occurs in traditional nonlinear optics when the signal beam and the
idler beam have the same frequency and the process of optical interference
is then denoted as {}``degenerate parametric oscillation''. A mechanical
version of the process was suggested in Ref.~\citep{fxue07-2}, where
the interference took place between two nanomechanical resonators
and it was shown to be the analog of a two-mode parametric down-conversion.

For our case, the procedure is half-optical (the polariton excitations)
and half-mechanical (the mirror excitations). To show the similar
squeezing effect in quadrature variance, we write the equations of
motion of the mirror operators from Eq.~(\ref{eq:quantized_ham})
\begin{eqnarray}
\dot{c} & = & -i(\Omega_{\mathrm{M}}+2\alpha_{\mathbf{n}})c-i2\alpha_{\mathbf{n}}c^{\dagger}-i\beta_{\mathbf{n}},\label{eq:eom}\\
\dot{c}^{\dagger} & = & i(\Omega_{\mathrm{M}}+2\alpha_{\mathbf{n}})c^{\dagger}+i2\alpha_{\mathbf{n}}c+i\beta_{\mathbf{n}}.\label{eq:eom_h.c.}
\end{eqnarray}
 The solution of the above equations, through Laplace transformation,
reads\begin{align}
c(t)= & \left[\cos\left(\Omega_{\mathrm{M},\mathbf{n}}t\right)-i\frac{\Omega_\mathrm{M}+2\alpha_{\mathbf{n}}}{\Omega_{\mathrm{M},\mathbf{n}}}\sin\left(\Omega_{\mathrm{M},\mathbf{n}}t\right)\right]c(0)\nonumber \\
 & -\left[\frac{i2\alpha_{\mathbf{n}}}{\Omega_{\mathrm{M},\mathbf{n}}}\sin\left(\Omega_{\mathrm{M},\mathbf{n}}t\right)\right]c^{\dagger}(0)\label{eq:soln_c}\\
 & +\frac{2\beta_{\mathbf{n}}}{\Omega_\mathrm{M}+4\alpha_{\mathbf{n}}}\sin^{2}\left(\frac{\Omega_{\mathrm{M},\mathbf{n}}t}{2}\right)-\frac{i\beta_{\mathbf{n}}}{\Omega_{\mathrm{M},\mathbf{n}}}\sin\left(\Omega_{\mathrm{M},\mathbf{n}}t\right).\nonumber \end{align}

We recognize that, unlike a typical optical parametric oscillator,
even when the mirror is set initially to a vacuum state $\left|0\right\rangle $,
the expectation value $\left\langle 0|c(t)|0\right\rangle $ will
be non-zero over time because of the perturbation from the polaritons.
As long as the numbers $n_{A}$ and $n_{B}$ of the polaritons are
not both zero at the same time, the inhomogeneous term $i\beta_{\mathbf{n}}$
on the right-hand-side of Eqs.~(\ref{eq:eom}-\ref{eq:eom_h.c.})
would become nonzero and the motion of the mirror would be initiated
by the incident polaritons, which is consistent with the vacuum state
entanglement we discussed in the last section. Compared to Ref.~\cite{fxue07-1}
for generating a squeezed entangled state of a mechanical resonator,
the requirement of preparing different initial Fock and coherent states
is lifted.

When the criterion Eq.~(\ref{eq:critical_cond}) is met, the variance
$\left\langle \Delta x^{2}\right\rangle $ in the coordinate quadrature
Eq.~(\ref{eq:cordinate}) demonstrates a squeezing effect
\begin{equation}
\left\langle \Delta x^{2}\right\rangle =\frac{2\cosh^{2}\left(\Omega_{\mathrm{M},\mathbf{n}}t\right)}{m\Omega_{\mathrm{M}}}+\frac{2\sinh^{2}\left(\Omega_{\mathrm{M},\mathbf{n}}t\right)}{m(\Omega_{\mathrm{M}}+4\alpha_{\mathbf{n}})}
\end{equation}
 where $\Omega_{\mathrm{M},\mathbf{n}}$ is meant in the equation
above to be the real magnitude of the pseudo-frequency $\Omega_{\mathrm{M},\mathbf{n}}$.

\section{\label{sec:conclusion}Conclusion and Remarks}

We have studied a cavity system composed of atoms, a cavity field
and a movable mirror and showed that the collective excitations of
the atoms are dressed by the cavity field and transformed into polaritons,
causing their entanglement with the cavity mirror. The mirror state,
in the adiabatic limit of the polaritons, is distinctly squeezed,
according to the number distribution of two polariton modes; and its
variance in coordinate space is also squeezed.

Before we conclude this paper, we remark a recent article by Paz and
Roncaglia~\citep{paz08} in which the entanglement dynamics between
two resonators at finite temperatures are classified into {}``sudden
death''~\citep{yu04}, {}``sudden death and revival'' and {}``no-sudden
death'' regions according to the amount of fluctuations the resonators
experience compared to their squeezing rate. Note that the squeezing
rate in our model Eq.~(\ref{eq:squeeze_opt}), defined through Eq.~(\ref{eq:bogo_tfm}),
is also related to the choice of operating regimes determined by Eq.~(\ref{eq:critical_cond}).
Therefore, we conclude that entanglement operates in regions of different
characteristics not only in finite temperature environments but also
in zero temperature settings, as shown by our model.

\begin{acknowledgments}
C.P.S. acknowledges supports by the NSFC with Grant Nos. 90503003
and the NFRPC with Grant Nos. 2006CB921205 and 2005CB724508. F.N.
gratefully acknowledges partial support from the National Security
Agency (NSA), Laboratory Physical Science (LPS), Army Research Office
(ARO), National Science Foundation (NSF) grant No. EIA-0130383, JSPS-RFBR
06-02-91200, and Core-to-Core (CTC) program supported by the Japan
Society for Promotion of Science (JSPS).
\end{acknowledgments}


\begin{thebibliography}{10}
\bibitem{mirror}Hereafter, any reference to the term {}``mirror''
refers solely and specifically to this movable mirror, which is of
our main interest here, not the fixed mirror at the opposite end of
the cavity.
\end{thebibliography}
\end{document}